\documentclass{aa}
\usepackage{graphicx,psfig}
\def\be{\begin{equation}}
\def\ee{\end{equation}}

\def\msun{\rm M_{\odot}}
\def\msunyr{\msun~{\rm yr^{-1}}}

\begin{document}
   \title{A Jet-ADAF model for Sgr A$^*$}


\author{Feng Yuan\thanks{Also at CAS-PKU Joint Beijing Astrophysical 
            Centre, Beijing 100871, China;
            and Astronomical and Astrophysical Center of East China,
            Nanjing University, Nanjing 210093, China.} 
        \and Sera Markoff\thanks{Humboldt Research Fellow} 
	\and Heino Falcke}

\offprints{F. Yuan; fyuan@mpifr-bonn.mpg.de}

\institute{Max-Planck-Institut f\"{u}r Radioastronomie, 
          Auf dem H\"{u}gel 69, D-53121 Bonn, Germany}

\date{Received 2001}


\abstract{
The recent {\em Chandra} observation of the radio source at the center
of our Galaxy, Sgr A$^*$, puts new constraints on its theoretical
models. The spectrum is very soft, and the source is rapidly variable.
We consider different models to explain the observations.
We find that the features of the x-ray spectrum
can be marginally explained with an
advection-dominated accretion flow (ADAF) model while it does not 
well fit the radio spectrum. 
An ADAF with strong winds (ADIOS) model 
is not favored if we assume
that the wind does not radiate. Alternatively, we propose a 
coupled jet plus accretion disk model to explain the observations for
Sgr A$^*$.  The accretion flow is described as an ADAF fed by
Bondi-Hoyle accretion of hot plasma in the Galactic Center region. A
small fraction of the accretion flow is ejected near the black hole,
forming a jet after passing through a shock.  As a result, the
electron temperature increases to $\sim 2 \times 10^{11}{\rm K}$,
which is about 10 times higher than the highest temperature attained
in the ADAF. The model is self-consistent since the main jet parameters
are determined by the underlying accretion disk
at the inner edge. The emergent spectrum of
Sgr A$^*$ is the sum of the emission from jet and underlying ADAF. 
The very strong Comptonization of synchrotron emission from the jet
dominates the bremsstrahlung from the ADAF, therefore, a very short
variability timescale is expected and the predicted X-ray slope and
the radio spectrum is in very
good agreement with the observations. 
}

\maketitle

\section{Introduction}

The energetic radio source Sgr A$^*$ located at the center of our
Galaxy is now widely believed to be the signature of a massive black
hole with mass $M=2.6 \times 10^6 \msun$ (Melia \& Falcke 2001; 
Haller et al. 1996; Eckart \& Genzel 1996; Ghez et al. 1998; Reid et
al. 1999; Backer \& Sramek 1999).  Its radio spectrum seems to consist
of two components, with a break around $\sim 50$ GHz. The spectral
dependence is $F_{\nu} \propto \nu^{0.2}$ for $\nu < 50 $ GHz, while
above this break there is a submm bump which is described by $F_{\nu}
\propto \nu^{0.8}$ up to $\sim 10^3$ GHz followed by a steep cut-off
towards the infrared (IR) (Zylka et al. 1992; Serabyn et al. 1997;
Falcke et al. 1998).  The upper limits from IR (Menten et
al. 1997) and ROSAT X-ray observations (Predehl \& Tr\"{u}mper 1994)
indicate that this source is quite dim. 

On the theoretical side, a number of models have been proposed in the
past years for Sgr A$^*$.  Most models are based on accretion onto the
central massive black hole.  Possible sources of accretion material include the
stellar winds emitted by the nearby massive stars and the hot
interstellar medium.  Since in either case the angular momentum of the
accretion flow should be small, Melia (1992; 1994) proposed a spherical
accretion model.  In this model the accretion flow is assumed to
free-fall until a Keplerian disk is formed within a small
``circularization'' radius.  The main contributors to the 
radio and X-ray spectra are 
synchrotron radiation and bremsstrahlung, respectively, from the
roughly free-fall flow beyond the small disk. However, spherical
accretion is likely to be an over-simplification, since the accretion flow
still possesses some angular momentum.  An advection-dominated accretion
flow (ADAF) model therefore is more dynamically exact in this sense 
(Narayan et al. 1995; Manmoto et al. 1997; Narayan
et al. 1998). The most attractive feature of 
the ADAF model is its ability to
explain the unusual low-luminosity of Sgr A$^*$ given the
relatively abundant accretion material.  This is because most of the
viscously dissipated energy is stored in the flow and advected beyond
the event horizon rather than radiated away (Ichimaru 1977; Rees et
al. 1982; Narayan \& Yi 1994, 1995; Abramowicz et al. 1995; Chen et
al. 1995; Narayan et al. 1997; Chen et al.
1997). In the application to Sgr A$^*$, the radio spectrum is produced
by the synchrotron process in the innermost region of the disk while
the X-rays are due to bremsstrahlung radiation of the thermal
electrons in a large range of radii 
$\sim 10^3-10^4 R_{\rm s}$, where $R_{\rm s} =2
GM/c^2$ is the Schwarzschild radius. However, the ADAF under-predicts the
low-frequency radio emission of Sgr A$^*$ by over an order of
magnitude and additional assumptions must be imposed in order to match
the spectrum (Mahadevan 1998, \"{O}zel et al. 2000).

Following the initial paper by Reynolds \& McKee (1980)
(see also Blandford \& K\"onigl 1979), Falcke et al. (1993)
proposed that it is the jet stemming from the disk rather than the
disk itself which is responsible for the radio spectrum of Sgr A$^*$. In this
model, the submm bump is produced by the acceleration zone of the jet, 
called nozzle, while the low-frequency radio spectrum comes from the part of
the jet beyond the nozzle (Falcke 1996b; Falcke \& Biermann 1999).  
The nozzle is of order 10$R_{\rm s}$ and forms
from the disk at a radius of $\sim 2 R_{\rm s}$. This model gives an excellent fit 
to the radio spectrum of Sgr A$^*$, including the low-frequency spectrum
below the break and the submm bump, but the expected X-ray emission
was not calculated explicitly.

The latest observational constraints for Sgr A$^*$ come from the
high spatial resolution ($\approx 1^{\arcsec}$) {\em Chandra X-ray
Observatory} (Baganoff et al. 2001a, 2001b). Baganoff et al. observed Sgr A$^*$
twice and they found that 
Sgr A$^*$ comes in two states: quiescent and flares. In the present
paper we concentrate on the quiescent state, whereas the flare state
is considered in Markoff et al. (2001b). 
The main observational results for the quiescent state
are summarized as follows\footnote{The luminosity and especially
the photon index are taken from Baganoff et al. 2001b, which are 
slightly different from those in Baganoff et al. 2001a where the spectral
models used did not account for dust scattering; see Baganoff et al. 2001b
for details.}.

\begin{itemize}
\item
The absorption-corrected 2-10 keV luminosity is 
(2.2$^{+0.4}_{-0.2}$)$\times10^{33}
{\rm erg ~s}^{-1}$.
\item The spectrum is well fitted by an absorbed
power-law model with photon index $\Gamma=2.2^{+0.5}_{-0.7}$.
\item The inner region of the source is rapidly variable on short
timescale of $\simeq 1 {\rm hr}$. A rapid drop of flux on a timescale of
10 minutes is detected in the flare state. On the other hand, the comparison 
between the two observations with an interval of about one year indicates
that the steady X-ray flux remains almost constant. 
\item Some fraction of the X-ray flux may come from a partly extended
region with diameter $\approx
1^{\arcsec}$. 
\item There is tentative evidence for a Fe K$\alpha$ line at 6.7 keV.
\end{itemize}
 
These results provide new and strict constraints to the 
theoretical models for Sgr A$^*$. In both the ADAF and
spherical accretion models mentioned above, the X-ray radiation is
produced by bremsstrahlung originating from $10^3$--$10^4R_{\rm s}$.
Hence the spectrum is very hard with photon index $\Gamma \sim 1.4$
and the predicted variability timescale is thousands of hours, much
longer than the observed $\sim 1$ hour. 

Therefore it is necessary to 
reexamine the theoretical models for Sgr A$^*$. 
Melia et al. (2001) proposed that the electrons in the
small Keplerian disk can attain a very high temperature through some
magnetic processes, and the resulting synchrotron and self-Compton
emission are responsible for the radio and X-ray spectrum.  However,
the formation of the small disk may not be a necessary result of such
low angular momentum accretion. An accretion flow with very low
angular momentum can still be described by an ADAF, although such
accretion may belong to the Bondi-like type rather than disk-like
type, as shown by Yuan (1999) (see also Abramowicz \& Zurek 1981;
Abramowicz 1998).  Thus the dynamical scenario of this model needs to
be studied carefully.  

For the jet model, Falcke \& Markoff (2000)
take into account the contribution from synchrotron self-Compton 
emission (SSC) in the nozzle and find
that the parameters required to interpret the submm bump give a
very good fit to the {\em Chandra} spectrum without changing the basic
parameters of the jet model. But the remaining important problem in the model is
why the parameters of the jet possess the required values,
particularly in reference to the inferred underlying accretion disk.
Previous ideas of a standard optically thick accretion disk in Sgr
A* (e.g., Falcke \& Heinrich 1994) do not seem to work 
because the predicted IR flux 
from a standard thin disk with a reasonable
accretion rate would be several orders of magnitude higher than the
observed IR upper limit (Falcke \& Melia 1997).
Therefore, it
is crucial to consider the jet and accretion flow as a coupled system
in Sgr A$^*$, and to consider what are their respective roles if both
are truly present in Sgr A$^*$.  Yuan (2000) presented the first
effort, by considering a combination of jet and ADAF models. However,
the complete {\em Chandra} data was not available at that time and the
detailed coupling mechanism was lacking in Yuan (2000) so it is
necessary to revisit the model again.
  
The development of the theory provides a new chance to model Sgr A$^*$.
Since the Bernoulli parameter of the ADAF is positive, which
means the gas can escape to infinity with positive energy, Blandford
\& Begelman (1999) propose an advection-dominated inflow-outflow
solution (ADIOS) in which most of the gas is lost through winds rather
than accreted past the horizon of the black hole. The concept of
strong winds from accretion flow was also proposed and studied by Xu
\& Chen (1997) and Das \& Chakrabarti (1999). The former described
pressure-driven winds from centrifugally supported
boundary layers and shocks in the inner regions of disks, and the latter 
proposed an advection-dominated flow where the central black hole
redirects the inward flow at low
latitudes into an outflow at high latitudes. We are not
explicitly making use of the latter two models. 
The most appealing point of the ADIOS model as applied to Sgr A$^*$ is that the
predicted X-ray spectrum is possibly much softer than that of the ADAF
(Quataert \& Narayan 1999), and therefore could possibly give a better fit to
the {\em Chandra} data. This is because the density profile of
the accretion flow becomes flatter due to the wind, while X-ray emission
at higher frequencies is produced in the inner region of the accretion
flow. If we assume that the mass accretion rate in the ADIOS is described
by a power-law of radius, $\dot{M}\propto R^{p}$, the predicted photon
index in {\em Chandra} band is approximately $\Gamma \approx 3/2+2p$.
Thus it is necessary to investigate this model for the possibility
of interpreting the {\em Chandra} results.

In this paper we explore several of the above-mentioned models
for Sgr A$^*$. By probing a larger parameter space than before, we
find that ADAFs can give a marginal  
interpretation to the new {\em Chandra} results, although the fit 
is not very good in some points (Sect. 2),
while the ADIOS model can't (Sect. 3). In Sect. 4
we propose that the combination of an ADAF and a jet could provide an
excellent fit to the observations to Sgr A$^*$, and present our model
results. The last section is a summary and discussion.

\section{ADAFs}

We first model Sgr A$^*$ with the advection-dominated accretion
model. The modeling technique is described in detail in Yuan et
al. (2000; see also Nakamura et al. 1997).  We use the Paczy\'nski \&
Wiita (1980) potential to mimic the geometry of the central black
hole.  A randomly oriented magnetic field is assumed to exist in the
accretion flow and the ratio between the gas pressure and total
pressure (gas pressure plus magnetic pressure) is denoted as $\beta$.
As commonly used, we assume that a fraction, $\delta$, of viscous
dissipation will directly heat electrons.  The radiation mechanisms we
consider include bremsstrahlung, synchrotron radiation and their
Comptonization.  We require a physical global solution that satisfies
the no-torque condition at the horizon of the black hole, a sonic
point condition, and the outer boundary conditions. The calculation of
the spectra and the structure of the accretion flows are made completely
self-consistent as the full set of coupled
radiation hydrodynamical accretion equations are solved numerically.

The parameters are adopted as follows. We take the black hole mass as
$M=2.5 \times 10^6 \msun$, and the viscosity parameter is fixed as
$\alpha=0.1$. We assume the magnetic field is in equipartition with
the gas pressure or weaker, i.e., $\beta =0.5, 0.9, 0.99$, although the
sub-equipartition magnetic field is more plausible if
$\alpha=0.1$. We set $\delta$ as $10^{-3}$ or $10^{-2}$
as usual, i.e., we assume that most of the
viscous dissipation will heat ions.  For the mass
accretion rate, using their latest {\em Chandra} observational data,
Baganoff et al. (2001a) estimate $\dot{M} \sim 3
\times 10^{-6} \msun~{\rm yr^{-1}}$, if the stellar wind is the
accretion material, or $\dot{M} \sim 1 \times 10^{-6} \msun~{\rm
yr^{-1}}$, if the hot ISM around Sgr A$^*$ serves as the accretion
source, which we use as our reference numbers. In principle, the
accretion rate of Sgr A* could be much higher, based on the available
material from stellar winds. However, explosive events like the
hyper-/supernova Sgr A East could temporarily reduce the accretion
rate onto Sgr A* substantially (e.g., Coker~2001).

The outer boundary conditions should be taken seriously since they may
affect the emergent spectrum significantly (Yuan et
al. 2000). Throughout this paper we set the outer boundary of
the accretion flow at $R_{\rm out}=10^5 R_{\rm s}$, where $R_{\rm s}$ is the
Schwarzschild radius of the black hole, since this is approximately
the location where the accretion begins according to the Bondi-Hoyle capture
theory.  Three outer boundary 
conditions are the temperatures of ions
and electrons, $T_{\rm i,e}$, and the angular velocity of the accretion
flow, $\Omega_{\rm out}$, at $R_{\rm out}$.  When $R_{\rm out}$ is very
large, as in the present case, the available range of $T_{\rm i,e}$
within which we can get a physical solution is small, therefore the
effect of $T_{\rm i,e}$ can be neglected. But the feasible range of
$\Omega_{\rm out}$ is large and may have a significant effect on the
emergent spectrum. For fixed parameters $R_{\rm out}$, $\beta$ and
$\alpha$, there exists a critical value of $\Omega_{\rm out}$ above
which the accretion is of disk-like type while below it 
is of Bondi-like type (Yuan 1999).  The density of the
Bondi-like accretion flow is much lower than the disk-like type at
the same mass accretion rate. Unfortunately the exact value of
$\Omega_{\rm out}$ is uncertain. We only know that it must be low no
matter whether it originates from stellar winds or from the hot ISM. For
example, the hydrodynamical simulations of Coker \& Melia (1997) found
$\Omega_{\rm out} \sim 0.2~
\Omega_{\rm Kepler}$, if it comes from stellar winds. We therefore
require in our model that $\Omega_{\rm out} \la 0.3~ \Omega_{\rm
Kepler}$.

\begin{figure}
\psfig{file=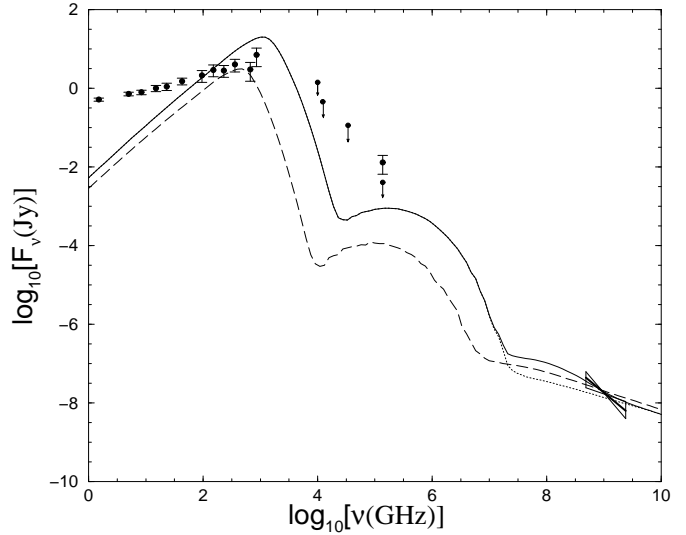,width=8.8cm,angle=270}
\caption{Two fits for a standard ADAF model of Sgr A$^*$.
The radio and IR data are compiled by Melia \& Falcke (2001). The short solid
line in the X-ray error box shows the best fit
to the {\em Chandra} observation by a power-law
model in Baganoff et al. (2001b).
The parameters for the solid line are
$\dot{M}=2.8 \times 10^{-5} \dot{M}_{\rm Edd}$, $\alpha=0.1, \beta=0.9,
\delta=10^{-2}$. For the dashed line, $\dot{M}=3 \times 10^{-5}
 \dot{M}_{\rm Edd}, \alpha=0.1, \beta=0.99, \delta=10^{-3}$.
The outer boundary conditions are $T_{\rm i}\approx
T_{\rm e}=8\times 10^6{\rm K}$ (both), $\Omega_{\rm out}=
0.16\Omega_{\rm Kepler}$ (solid line) and $\Omega_{\rm out}=
0.15\Omega_{\rm Kepler}$ (dashed line).}
\end{figure} 

Fig. 1 shows our fits to the spectrum with an ADAF model. The
radio and IR data are compiled by Melia \& Falcke (2001). The short 
solid line within the error box shows the best fit 
to the {\em Chandra} observation by a
power-law model in Baganoff et al. (2001b). We show $F_{\nu}$ rather
than $\nu F_{\nu}$ because it is more conducive for judging the
quality of the spectral fit at radio bands. The solid line
shows our best fit to the spectrum especially to the {\em Chandra data}.
The parameters are
$\alpha=0.1, \beta=0.9, \delta=10^{-2}$, and $\dot{M}= 1.55 \times
10^{-6}\msun~{\rm yr}^{-1}$ (or $ 2.8 \times 10^{-5}\dot{M}_{\rm
Edd}$,  here the Eddington accretion rate is
defined as $\dot{M}_{\rm
Edd}= L_{\rm Edd}/0.1c^2= 5.525 \times 10^{-2} \msun~{\rm yr}^{-1}$
where $L_{\rm Edd}$ is the Eddington luminosity.).  The outer boundary
condition at $R_{\rm out}=10^5R_{\rm s}$ is $T_{\rm i}\approx T_{\rm e}=8
\times 10^6{\rm K}$ (this value of temperature 
is consistent with the X-ray observations by
Baganoff et al. 2001a), $\Omega_{\rm out} =0.16\Omega_{\rm Kepler}$.

From the figure we find that an ADAF model can fit the X-ray spectrum  
with a reasonable accretion rate \footnote {In
Yuan (2000), the {\em Chandra} flux is produced with a higher
accretion rate, $1.5 \times 10^{-4}\dot{M}_{\rm Edd}$. The discrepancy
in accretion rate is because in Yuan (2000) the accretion is
Bondi-like, while in the present paper it is disk-like. For the same
accretion rate, the density in a Bondi-like accretion flow is much
lower than in a disk-like flow. Since bremsstrahlung emission is
proportional to the square of the density the X-ray emission can be
largely different between the two cases.} although the predicted spectrum is 
flatter than the best fit of Baganoff et al. (2001b).
The predicted X-ray spectrum is
composed of two components, namely the bremsstrahlung from the outer region
of ADAF and the second-order SSC from
the innermost region of ADAF. The dotted line in the figure shows
the result excluding the SSC component. Different from our result, 
in the ADAF model of Narayan 
et al. (1998), the X-ray emission is dominated by
bremsstrahlung alone. One reason for the difference is our use of a higher
$\delta$, and another reason is that we treat the outer
boundary conditions more carefully. For bremsstrahlung, 
the emission at a frequency $\nu$ is dominated by the largest radius in
an ADAF that satisfies $h\nu \sim kT(r)$. Our numerical calculation
results indicate that the 2 and 10 keV radiation is dominated by radii around
$4\times 10^4 R_{\rm s}$ and $7400R_{\rm s}$, respectively. This large radial
 range is consistent with the extended emission component
($\approx 1^{\arcsec} \approx 10^5 R_{\rm s}$) observed by {\em
Chandra}. In addition, the thermal bremsstrahlung can also explain the possible 
Fe K$\alpha$ emission line at 6.7 keV (Narayan \& Raymond 1999).
The dynamical timescale of the accretion flow at these large radii, which is
responsible for the bremsstrahlung variability, is $t_{\rm d}\approx
(R^3/GM)^{1/2} \sim 1$ year. 
Baganoff et al. (2001b) made a comparison between their two 
observations with an interval
of about one year and found that the steady state
X-ray flux remains almost constant. This result,
combined with the rapid variability, seems to indicate
that there are two components to the X-ray emission operating
on very different spatial scales and having very different time scales
for variability. Bremsstrahlung may well be the component responsible for the
constant flux. The SSC component mainly
comes from regions $\la 3 R_{\rm s}$ (see Fig. 1 in Manmoto et al. 1997).
The corresponding variability timescale is $\sim 3R_{\rm s}/v_r \approx 1000$ seconds.
So this component would be responsible for the observed rapid
variability. 

However, as shown by Fig. 1, this model over-predicts 
the submm bump by a factor of $\sim$ 2-3. We then try to 
lower the synchrotron flux from the 
ADAF to fit the submm bump better, as shown by the dashed line in Fig. 1.
The parameters are $\dot{M}= 1.66 \times
10^{-6}\msun~{\rm yr}^{-1}, \beta=0.99, \delta=10^{-3}$. The outer 
boundary conditions are the same as the solid line except with
 $\Omega_{\rm out}=
0.15\Omega_{\rm Kepler}$.
In this case the second order SSC will become
too weak to contribute to the X-ray flux, therefore,
bremsstrahlung is almost the sole contributor to the X-ray spectrum. 
Consequently, the predicted spectrum is too flat and
the $\sim 1$ hour variability is hard to explain. 
Considering that we can only investigate a limited parameter space of the ADAF
model, and the fact that the solid line only fits marginally, we conclude that
it is possible to interpret the spectrum of Sgr A* from submm
bump to X-ray using the ADAF model. 
However, it remains to be seen whether the
current ADAF model can indeed produce a strong flare as found by Baganoff
et al. (2001b). In addition, as in all previous ADAF models in the literature,
the ADAF model always under-predicts
the low-frequency radio spectrum which
needs a contribution from another component such as a jet. 

\section{ADIOS}

We next attempt to model Sgr A$^*$ with an ADIOS.  The modeling
approach is exactly the same as with the ADAF, except that the
accretion rate is assumed to be described by
$\dot{M}=\dot{M}_0(R/R_{\rm out})^p$.  We solve the full set of
coupled radiation hydrodynamical accretion equations to obtain the
spectra and the structures of the accretion flow consistently.  Note
that this is an improvement compared to Quataert \& Narayan (1999)
where some dynamical quantities such as radial velocity and sound
speed obtained in corresponding ADAFs (with $\dot{M}= \dot{M}_0$) are
used in calculating the spectra of the ADIOS. Following Quataert \&
Narayan (1999), we assume that the wind does not radiate.

We first assume that the fraction of viscous heating of electrons is
$\delta=10^{-3}$. We set $\alpha=0.1$ but treat 
$\dot{M},~ p$ and $\beta$ as
free in order to find the best set of parameters to 
fit the submm bump and the X-ray spectrum. The dashed line 
in Fig. 2 shows our best
model results. The parameters are $\dot{M}_0= 1.66 \times 10^{-5}
\msunyr$, $p=0.28$, $\alpha=0.1$, but $\beta=0.5$ (not 0.9
since otherwise the predicted radio flux is too low compared to the
observation). The outer boundary conditions are $T_{\rm i} \approx
T_{\rm e} =8 \times 10^6{\rm K}$, and $\Omega_{\rm out}=0.295
\Omega_{\rm Kepler}$ at $R_{\rm out}=10^5R_{\rm s}$. Compared to the ADAF model,
both the slope of the X-ray spectrum and the submm bump are now fitted better.
However, there are two serious problems for this fit. The first one is that the
required mass accretion rate is over 5 times higher than the upper limit
estimated in Baganoff et al. (2001a) mentioned above. The second problem
is that the X-ray spectra are produced by thermal bremsstrahlung
emission alone, therefore this model cannot explain the short timescale
variability. In fact, the introduction of a wind makes the
variability timescale even longer because the decreasing density of
accretion flows (e.g. Di Matteo et al. 2000) makes things worse.

The very rapid variability observed by {\em Chandra} indicates that
the X-ray emission comes from a very small spatial region. This points
towards SSC occurring in the inner
region of the disk. In the case of the existence of strong winds, the
density of the accretion flow in the innermost region is very low.
When the flow is tenuous, SSC will show some spectral peaks
as a result of different scattering orders.  To make SSC dominate over
bremsstrahlung in the X-ray band, the first order of SSC is more
promising due to the rapid decrease of Compton scattering probability
with increasing scattering orders. To make the first order SSC
component reach the {\em Chandra} band, the electron temperature in
the emission region must be very high. 

An effective way to increase the electron temperature in the accretion
flow is to increase $\delta$. In the ADAF we generally assume that
$\delta$ is as as small as $\delta=10^{-3}$ or $10^{-2}$, i.e., the viscous
dissipation mainly heats the ions.  However, because of the
uncertainty in the microphysics of the ADAF, it is possible that for
some reasons, such as magnetic reconnection, the viscous dissipation may
prefer heating electrons, i.e., $\delta$ may be much larger
(Bisnovatyi-Kogan \& Lovelace 1997, 2000; Gruzinov 1998; Quataert \&
Gruzinov 1999; Blackman 1999). In this case, the temperature of the
electrons will be greatly increased.
                                         
\begin{figure}
\psfig{file=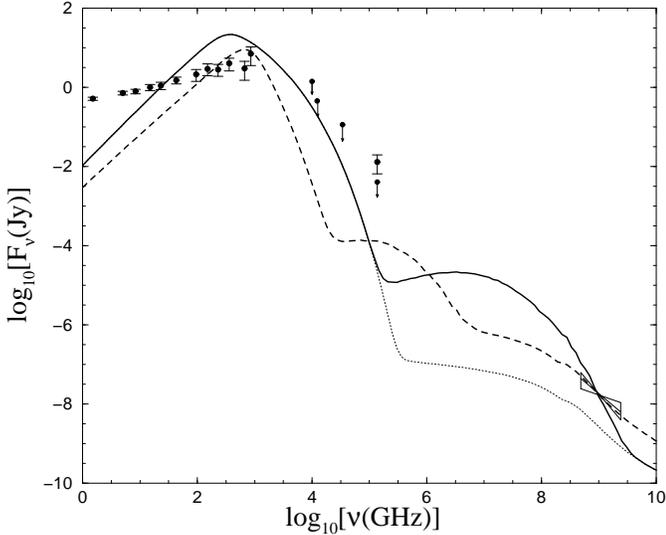,width=8.8cm,angle=270}
\caption{Three ADIOS spectral models for Sgr A$^*$.
The short-dashed line is for
$\delta=10^{-3}, p=0.28$, the long-dashed line for
$\delta=10^{-3}, p=0.6$. The  solid line is for
$\delta=1, p=0.4$, the dotted line
is exactly the same model as the solid line,
except that Comptonization of synchrotron radiation
is neglected. See text for other
parameters. }
\end{figure}

We try to model the spectrum using various values for $\delta$.  We
find that only when $\delta \approx 1$, i.e., almost all of the
viscous dissipation heats only electrons, can we get a high enough
electron temperature to make the first order SSC dominate the X-ray
emission.  The solid line in Fig. 2 shows such an example.  Other
parameters in this model are $\dot{M}_0=2.8 \times 10^{-6} \msunyr,
\alpha=0.1, \beta=0.9$, and $p=0.4$. The outer boundary conditions are
$T_{\rm i,e}=10^7{\rm K}$, $\Omega_{\rm out}=0.25\Omega_{\rm Kepler}$. The
temperature of electrons is as high as $10^{11}$~K for the accretion
flow within $ 6R_{\rm s}$ and the highest temperature is $3 \times
10^{11}$~K. This model is then very similar to the model proposed by Melia
et al. (2001) for Sgr A$^*$ in the sense that a high-temperature inner disk 
forms, with $T_{\rm e} >10^{11}$K. Synchrotron emission 
in this hottest region produces the
submm bump, synchrotron self-Compton dominates the X-ray band and
gives a very soft spectrum. The thermal bremsstrahlung radiation only
contributes a small part as shown by the dotted line, where SSC is
neglected. In this case a very short X-ray variability timescale can
be expected.

Putting aside the reality of such a high $\delta$, the fit is not
satisfactory on the following points: First, it under-predicts
the low-frequency radio spectrum.  Second, the predicted X-ray slope is
much steeper than the best fit of Baganoff et al. (2001b).
The third problem is that this
model over-predicts the flux above $\sim 100$ GHz by a factor of 4-6.
We cannot get a better fit no matter how we adjust the
parameters. Because of the strong self-absorption of synchrotron
emission, the radio spectrum is the result of a super-position of
blackbody radiation from the different parts of the ADAF with
different temperatures. So, comparing this model with an ADAF (or ADIOS
with small $\delta$), we can understand that the main reason for the
over-prediction is its too extreme temperature making the flux
of the blackbody radiation stronger.  Thus we conclude that, if we do
not consider the possible radiation of winds, the ADIOS model is not
favored for Sgr A$^*$.

However, the approximation that the wind does not radiate may be an
over-simplification.  For example, the part of the wind originating
from the supersonic region of the accretion disk will possibly be
shocked when it is ejected out of the disk. Thus it would reach very
high temperatures and its radiation could not be neglected. In this
sense, the wind within the sonic radius will present itself as
radiative, outflowing plasma--i.e., like the plasma jets
typically observed in AGN. The model would then possibly become 
similar to our jet-disk model presented below.

\section{Jet-ADAF model for Sgr A$^*$}

The idea of combining a jet and an ADAF was proposed by Falcke (1999)
and Donea et al. (1999).  Yuan (2000) first worked this
out in detail and calculated the spectrum of the jet-ADAF system for
Sgr A$^*$ and some nearby elliptical galaxies. There is only scant
direct observational evidence for the existence of a jet in Sgr A$^*$,
from the near-simultaneous VLBA measurements by Lo et al. (1998). They
found that the intrinsic source structure at 43 GHz is elongated along
an essentially north-south direction, with an axial ratio of less than
0.3. However, it is interesting to note that the nearby spiral galaxy
M81 has a very similar radio core and similar unusual polarization
features as Sgr A$^*$ (Bietenholz et al. 2000; Brunthaler
et al. 2001).  In this source, a jet was clearly observed after many
VLBI observations, with the length of the jet being only $\sim 400$AU at 43
GHz (Bietenholz, Bartel, \& Rupen 2000). If we consider M81 to be a
scaled-up version of Sgr A$^*$, as suggested by their similarity, there
could well exist a jet in Sgr A$^*$ as well.  Of course, the jet in
Sgr A$^*$ would be less powerful and hence smaller, making it difficult
to detect because of the strong scattering of radio waves within the
Galaxy.  More generally, jets seem to be symbiotic with accretion
disks (Falcke \& Biermann 1995; Livio 1999) and they are found in
basically all kinds of accretion powered systems. In this sense the model
presented here may be quite general. 

The picture of our jet-disk model presented here is as follows.  The
accretion disk is described by an ADAF.  In the innermost region, $r <
r_0$, where parameter $r_0$ is the jet location, 
a fraction $q_{\rm m}$ of the accretion flow is ejected out of
the disk and forms a jet. Since in our model $r_0$ is very small ($r_0
\approx 2 R_{\rm s}$, within the sonic point of the accretion disk), 
the radial velocity of the accretion flow
is supersonic at this small radius (the Mach Number is $\sim
2-3$). Therefore, when the supersonic accretion flow is transferred
from the disk into the jet, which is normal to the disk,
the plasma will be shocked before entering into the jet.  The shocked gas
passes through a nozzle where it becomes supersonic.  Then it is
accelerated along the jet axis through the gas pressure gradient force
(the gravitational force is ignored since its effect is rather small
in the supersonic regime far away from the black hole) and expands
sideways with its initial sound speed.  Given the initial physical
states of the plasma at the sonic point (top of the nozzle), we can
solve for all the quantities as a function of distance from the nozzle,
and after calculating the density and 
the strength of the magnetic field, we can calculate 
the radiation of the jet (Falcke \& Markoff 2000). 

If, however, there exists a possibility that a substantial fraction of
the accretion flow can be transferred into the jet directly without
being shocked (e.g., the accretion flow outside of the sonic point
also goes into the jet), 
we could also envisage a mixture of a relatively cold
(un-shocked, $\sim 10^{10}$K in the innermost 
region of ADAF) and hot (shocked, $\sim 10^{11}$K, see below
for this value) electrons in the jet.
If the energy transfer between 
particles is inefficient, this kind of mixture
could last for a long distance along the jet. For an emission model we
can ignore this possibility, because the implied radiation should be
less than the dashed line in Fig. 3 and the contribution to the
overall spectrum can be neglected. On the other hand, such a mixture
of hot and cold electrons may be needed when considering the circular
polarization of Sgr A$^*$ (Beckert \& Falcke 2001). This might
increase the coupling constant between jet and disk.

The exact physics of the nozzle are
difficult to model since we are at present unclear as to the physical
mechanism of jet formation. In this paper we treat the nozzle only
phenomenologically when calculating its spectrum.  We simply assume
that it consists of a series of cylinders with the same electron
temperature but linearly decreasing density (increasing velocity) from
bottom to top. The velocity of the gas at the base of the nozzle is
assumed to be $1/5$ of that at the top of the nozzle where it reaches
sound speed. The emission is not very sensitive to the exact value of
the initial nozzle speed. The main radiation mechanisms are
synchrotron emission and its Comptonization.  The parameters describing the jet
include radius and height of the nozzle, $r_0$ and $z_0$, 
electron temperature $T_{\rm e}$, electron number density $n_e$, the strength of
the magnetic field $B$ at the top of nozzle, and the angle between the
jet axis and the line of sight $\theta$.

All above are free parameters in the original jet model (but most of
them have obvious physical constraints to their range of values, see
Falcke \& Markoff 2000). But here in our coupled jet-disk system, more
constraints are required so that the jet parameters are consistent
with the underlying accretion disk.  $T_{\rm e}$ is calculated
self-consistently as follows. When some accretion gas passes through
the shock and enters into the jet, the ordered kinetic energy in the
pre-shock gas will be converted into thermal energy in the shock
front. Neglecting the effects of the magnetic field on the shock 
jump condition (we will check the rationality of this 
approximation later), we 
calculate the electron temperature of the post-shock plasma
by the following Rankine-Hugoniot relations, 
namely the conservation of flux
of mass, momentum, and energy. 
Written in the conventional notation, they are
\be 
[\rho v]=0,
\ee
\be
[P+\rho v^2]=0,
\ee
\be
\left[\frac{1}{2}v^2+\frac{2}{5}\frac{P}{\rho}\right]=0,
\ee
respectively. To obtain $T_{\rm e}$ immediately 
after the shock, we still need the ratio between
the ion and electron temperatures, $\xi$.
On the one hand, shock heating may favor the ions rather than electrons,
as isotropization of the bulk flow velocities will give to 
each species a thermal energy proportional to its mass. 
On the other hand, Coulomb collisions, and maybe collisionless processes also,
will bring about equilibration between ions and electrons temperatures.
Determining the value of $\xi$ is a difficult task (see
Laming 2000 for a recent review). In the context of supernova remnant
shocks,  Cargill \& Papadopoulos (1988) predict $\xi=5$ from
a numerical simulation, while Laming et al. (1996) derive
$\xi=20$ from their fit to observations. 
We set $\xi=10$ in our model.

For a given shock location $r_0$, we first  solve the
radiation-hydrodynamical equations describing the underlying ADAF
under selected parameters and outer boundary conditions to obtain the
pre-shock physical quantities at $r_0$.
We then substitute them in the above shock relations to get the
post-shock values. Thus we obtain the electron
temperature $T_{\rm e}$ in the nozzle.  Therefore $T_{\rm e}$ is not a free
parameter in our model, we can change its value only through changing
the parameters and outer boundary conditions of the underlying
accretion disk.  We use the same ``magnetic parameter'' $\beta$ as in
the ADAF to describe the ratio between the gas pressure and total
pressure in the nozzle to obtain the value of $B$ if
temperature and density are known. This means that $B$ is no longer a
free parameter, either. The density, $n_{\rm e}$, in the jet follows
from the coupling constant $q_{\rm m}$ in the jet-disk symbiosis model
(Falcke et al. 1993). This is defined as the ratio
between mass loss in the jet and accretion rate outside $r_0$. 
For jets, $q_{\rm m}$
is typically a few percent and we require $q_{\rm m} \ll
1$.

\begin{figure}
\psfig{file=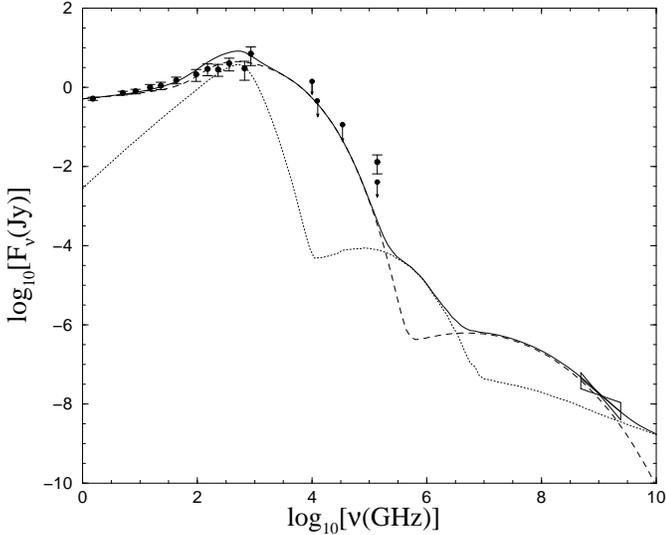,width=8.8cm,angle=270}
\caption{The jet-disk spectral model for Sgr A$^*$.
The dotted line is for the ADAF contribution, the dashed line is for
the jet emission, and the solid line shows their sum. See text for
details.}
\end{figure}

Our best spectral fit is presented by the solid line in Fig. 3. The
dashed line denotes the emergent spectrum from the jet, and the dotted
line is from the underlying disk (ADAF).  The solid line is their
sum. For ADAFs, the parameters are $\dot{M}=8.8 \times 10^{-7}~
\msunyr$, $\alpha=0.1, ~\beta=0.95$ and $\delta=10^{-3}$. The outer
boundary conditions are $T_{\rm i}\approx T_{\rm e}=8\times 10^6~{\rm K}$,
$\Omega_{\rm out}=0.27~\Omega_{\rm Kepler}$ at $10^5~ R_{\rm s}$. The
parameters for the jet are $r_0=1.7~R_{\rm s}$, $z_0=3.5~r_0$, $q_{\rm
m}=~0.5\%$ and $\theta=~35^{\degr}$, the ``calculated parameters'' are
$B=23{\rm G}$, $T_{\rm e}=2.1 \times 10^{11}~{\rm K}$, and $n_e=~2.4 \times
10^6$.  The mass loss rate in the jet is $\dot{M}_{\rm jet}= \pi r_0^2
c_s n_e m_p =4.3 \times 10^{-9}~\msun~{\rm yr}^{-1}$, i.e. 0.5\% of the
accretion rate. For $\beta =0.95$ and shock location $r_0=1.7R_{\rm s}$, 
the Alfv\'en Mach number $M_A \equiv v/v_{A1} =
(v/c_s)(4 \pi \gamma p/B^2)^{1/2} \approx 10 > 6$,
here $v_{A1} \equiv B_1/(4\pi\rho_1)^{1/2}$ is the pre-shock Alfv\'en speed,
the magnetic effects are weak in the shock
transition condition, so our hydrodynamic approximation to the shock 
transition condition, Eqs. (1)-(3) above, is justified (Draine \& McKee 1993).

This model fits the spectrum over the whole range of frequencies
from radio to the X-ray
quite well.  The submm bump is slightly over-predicted, but it is
acceptable considering the variability of the data in this band (Melia
\& Falcke 2001) and the uncertainty of the model.  The low-frequency
radio emission is mainly contributed by the jet outside the nozzle. The
contribution from the ADAF is rather weak and can be neglected.
The submm bump is the sum of the synchrotron
radiation from both the ADAF and the nozzle of the jet. We note that
the emission from the nozzle is much weaker than in the ADIOS with
$\delta=1$ presented in the last section (solid line in Fig. 2)
although the electron temperatures are both $\ga 10^{11}{\rm K}$.
Such a difference is not surprising considering the much
smaller spatial scale of the nozzle, $r_0=1.75R_{\rm s}$, while in that
case, there is a larger radial range with high temperature.  In this
sense, an abrupt increase in the temperature profile is necessary to
model the spectrum. This is naturally satisfied in our jet-disk model
by the formation of a jet. If instead the nozzle in our model
is replaced by a similar high-temperature component
such as the inner region of a disk, since the
temperature profile of the disk is in general smooth, the radial
range of this high-temperature component would be considerable. 
In this case, we expect
that the model would greatly over-predict the submm flux, and the low-frequency
radio spectrum is still hard to explain, as
indicated by our calculation for the ADIOS model with 
high $\delta$ in Sect. 3. This is the reason why in Melia et al. (2001)'s 
model the authors require an accretion disk as small as 
$\sim 5 R_{\rm s}$.  
 
The X-rays are mainly produced in the nozzle by SSC,
although bremsstrahlung from the ADAF also contributes
to a small degree.  The predicted X-ray spectrum is the sum of the
very soft SSC from the nozzle and the relatively flatter bremsstrahlung
spectrum from the ADAF, which is in very good agreement with the
{\em Chandra} data, almost identical to the best fit of Baganoff et
al. (2001b).  The fit is also much better than that of the ADIOS with
$\delta=1$ in the last section. In both cases, the X-ray emission is
the sum of bremsstrahlung and SSC, but in the present case,
bremsstrahlung produces a much flatter spectrum than in the case of an ADIOS
due to the absence of a strong wind.  Because of the contribution of SSC
from the jet, the variability timescale of X-rays should be short, $ t
\approx r_0/v_{\rm jet} \approx 10$ minutes. This 
is consistent with the $\sim$ 1 hour variability timescale determined
in the quiescent state and is in excellent
agreement with the $600$ seconds variability timescale detected
in the flare state. We show that it is the variability of the flux
from the nozzle that causes the huge-amplitude flare (Markoff et al. 2001b).
On the other hand,
since the bremsstrahlung radiation from the ADAF also contributes
partly to the X-ray spectrum, this could explain the possibly detected
extended source with $\sim 10^5 R_{\rm s} (\approx 1^{\arcsec})$, the 6.7 keV
K$\alpha$ emission line, and steady X-ray flux on $\sim $ one year timescale,
as we stated in Sect. 2.
                                                   
We note that the above nozzle parameters, temperature, spatial size
and density, are very close to the ``second component'' in the model
of Beckert \& Duschl (1997), which is also responsible for the
submm bump of Sgr A$^*$. These parameters seem to be the best ones to
fit the submm bump.   It is interesting that the nozzle with
these parameters will ``evolve'' naturally into a jet whose emission
can well reproduce the low-frequency radio spectrum of Sgr A$^*$, and
the Comptonization of its synchrotron emission can produce a very soft
X-ray spectrum which can fit the {\em Chandra} data excellently.  In
fact, to make the up-scattered submm bump extend to the {\em Chandra}
band, an electron temperature as high as $10^{11}{\rm K}$ is
needed. The peak frequency of this bump is $\sim 10^{12}$ Hz. To
up-scatter it to the X-ray band, $\nu \sim 10^{16}$ Hz, the electron
Lorentz factor must satisfy $4\gamma_e^2 \approx 10^{16}/ 10^{12}
\approx 10^4$.  This corresponds to a temperature of $T \approx
\frac{1}{k}\frac{\gamma_e}{3.5}m_ec^2 \approx 10^{11}$K.  This value
is about 10 times higher than the highest temperature that a canonical
ADAF can reach in its innermost region, but is naturally reached when
some fraction of accretion matter is shocked\footnote{We note in this context
that Falcke (1996b) and Beckert \& Duschl (1997) demanded that $T
\approx 10^{11}$K based on the Sgr A* spectrum alone.}. In addition to
a high temperature, the spatial size of the dominant emission medium
must be small, otherwise the model will over-predict the high-frequency
radio flux as in ADIOS with high $\delta$ (the solid line in Fig.
2). This is also easily satisfied in the jet model by requiring a
small $r_0$.  In addition to the above parameters, a truncated (no
hard tail) electron energy distribution is also required in the model,
otherwise the synchrotron emission will extend above the observed IR
flux upper limit.  Beckert \& Duschl (1997) simply assume a
mono-energetic distribution.  In our model, a relativistic thermal
distribution, which is highly peaked at $\gamma \approx
3.5\frac{kT}{m_ec^2}$, is a natural result of shock heating (e.g. Drury 1983)
since the Mach number is not very large in our case, $\sim 2-3$.

The mass accretion rate of the ADAF in our model, $8.8 \times
10^{-7}\msunyr$, is only marginally smaller than the lower limit of
Baganoff et al. (2001a) estimate of $1 \times 10^{-6}\msunyr$.  If we
used a higher accretion rate, we would obviously slightly over-predict
the flux at the submm bump band because of the higher flux of the
synchrotron emission from the ADAF.  

There are various ways to further evolve the model.  One is to
introduce global winds in the ADAF. The X-ray radiation from the disk
would be almost unaffected but the radio emission from the disk
would be greatly decreased because of the great decrease in
density close to the black hole (Quataert \&
Narayan 1999). But the wind cannot be too strong, otherwise the X-ray
spectrum would be too soft, as we argued in the case of ADIOS model
with $\delta \sim 1$.  Another modification is to assume that the
accretion disk is radiatively truncated within $r_0$, the radius of
the jet formation (Yuan 2000). The physical reason for the truncation
is that to form a jet, some amount of energy is needed.
If we assume this energy comes from the underlying disk, the
disk will be left cold within $r_0$ because of the energy extraction
(Blandford \& Payne 1982).  This will greatly suppress the 
synchrotron emission due to the very sensitive dependence of synchrotron
radiation on the temperature.

\section{Summary and Discussion}

Recent {\em Chandra} X-ray observations put new constraints on 
the theoretical models of Sgr A$^*$. The spectrum is very soft, the flux
is rapidly variable and the source is extended.
In this paper we consider three different models to explain the
observational results of Sgr A$^*$. 
We find that an ADAF model can give a marginally satisfactory interpretation to 
the {\em Chandra} spectrum and the rapid X-ray
variability. But our best fit is still not good for the radio spectrum
in the sense that it over-predicts the high-frequency radio
by a factor of 2-3 and significantly under-predicts the low-frequency radio.
We then consider the possibility of strong winds from 
ADAFs, i.e., an ADIOS model. If the winds are
non-radiative and viscous dissipation in the accretion flow mainly
heats ions, as generally assumed in the literature, this model
can fit the spectrum ranging from submm bump
to X-ray quite well. However, it is hard to explain the 
rapid X-ray variability
since in this model bremsstrahlung is the sole contributor at X-ray
band. If we assume that
most of the viscous dissipation preferentially heats electrons, 
a rapidly variable X-ray spectrum is expected since in
this case the X-ray emission is dominated by SSC.
But in this case the model over-predicts the radio flux above
$\sim 100$ GHz by a factor of 4-6, and the predicted X-ray spectrum is much
steeper than the best fit of the {\em Chandra} observations.

An excellent fit to all the data including low-frequency radio can be obtained 
with a coupled jet-disk model. In
this model, the accretion disk is described by an ADAF.
In the innermost region of the ADAF, $\sim
2R_{\rm s}$, some fraction $q_{\rm m}$ ($\sim 0.5\%$ if any
cold jet component is neglected. See our discussion in Sect. 4 for the
possibility of a cold jet component) of the accretion flow 
is ejected out of the ADAF
and transferred into the jet. In this process, a shock occurs because
the accretion flow is radially supersonic before the shock. After the shock the
temperature of electrons in the nozzle (the base of the jet) reaches
about $2 \times 10^{11}{\rm K}$. In this case, 
the synchrotron emission in the
nozzle largely dominates the submm bump, and its Comptonization
dominates the quiescent X-ray spectrum in Sgr A$^*$. The X-ray
spectrum is soft and the variability timescale is short. Out of the nozzle,
the jet gas expands freely outward under the force of the 
gas pressure gradient of gas pressure.
Furthermore its self-absorbed synchrotron radiation
gives an good fit to the low-frequency radio spectrum of Sgr A$^*$ which 
is hard to explain in ADAF models. The model is completely self-consistent.

The jet in our model produces a slightly inverted radio spectrum, as
can be understood from the canonical model of Blandford \& K\"onigl
(1979), with modifications as in Falcke (1996a).  In the absence of a
shock acceleration region in the highly-supersonic outer region of the
jet, the particles retain the highly-peaked relativistic Maxwellian
energy distribution which is attained by shock {\it heating} occurring
when the radial supersonic accretion flow is transferred into the
vertical direction.  On the other hand, the electrons in AGN jets
typically seem to have a power-law high-energy tail after shock acceleration
in jets, since the Mach number in jets is very high (Drury 1983).
In that case, a corresponding optically thin power-law spectrum at
IR/optical frequencies is generally expected, as is seen in many AGN
and perhaps even X-ray binary jets (e.g., Markoff et al.
2001a). In the case of Sgr A$^*$, the absence of an optically-thin
power-law indicates that, for some unknown reason, 
such high Mach number shocks do not occur.
If they would occur under certain conditions,
we should still see an inverted radio spectrum, but we would also expect
some kind of hard power-law emission at higher frequencies (mid-IR to
X-rays).

In addition to the observations we mention in the present paper, there
are also constraints to the model through the frequency-size
relationship obtained from VLBI observations (Rogers et al. 1994;
Krichbaum et al. 1998; Lo et al. 1998).  The jet-disk model can fit
this well as shown in Falcke \& Markoff (2000).

We therefore conclude that it is possible to present a consistent
picture of the emission processes associated with the central black
hole in our Galaxy by combining the three basic astrophysical
ingredients that have been discussed in recent years: Bondi-Hoyle
accretion from the immediate environment, optically thin accretion
through an ADAF, and energy extraction and visible emission by a
plasma jet. Our jet-ADAF model predicts a closely correlated
variability among sub-millimeter, IR, and X-ray. 
More broad-band observations and monitoring at various
wavebands (radio, IR, X-rays) 
will help to judge whether it will be
possible to establish a standard model invoking those elements
for Sgr A* in the near
future. For example, more precise determination of the IR
flux will help to further discriminate between the jet-ADAF model and
the pure ADAF model since the former predicts higher IR flux than the latter. 
This will also be crucial for understanding the activity in
low-power black holes in general.

\begin{acknowledgements}
We are grateful to Peter Biermann for discussions on shock physics.
F. Y. thanks the partial support from China 973 Project under
NKBRSF G19990754.
\end{acknowledgements}

\end{document}